# Scientific production in the era of Large Language Models

Keigo Kusumegi[1]†, Xinyu Yang[1]†, Paul Ginsparg[1], Mathijs de Vaan[2]*, Toby Stuart[2]*, Yian Yin[1]*

[1]Department of Information Science, Cornell University, Ithaca NY, USA

[2]Haas School of Business, University of California Berkeley, Berkeley CA, USA

†These authors contributed equally to this work

*Correspondence to: mdevaan@haas.berkeley.edu, tstuart@haas.berkeley.edu, yian.yin@cornell.edu.

**Large Language Models (LLMs) are rapidly reshaping scientific research. We analyze these changes in multiple, large-scale datasets with 2.1M preprints, 28K peer review reports, and 246M online accesses to scientific documents. We find: 1) scientists adopting LLMs to draft manuscripts demonstrate a large increase in paper production, ranging from 23.7-89.3% depending on scientific field and author background, 2) LLM use has reversed the relationship between writing complexity and paper quality, leading to an influx of manuscripts that are linguistically complex but substantively underwhelming, and 3) LLM adopters access and cite more diverse prior work, including books and younger, less-cited documents. These findings highlight a stunning shift in scientific production that will likely require a change in how journals, funding agencies, and tenure committees evaluate scientific works.**

The scientific enterprise is intimately connected with technological innovation. The microscope (*1*), advances in computing (*2, 3*), and next-generation sequencers (*4*), for example, shifted the frontier of research. Today, the fast adoption of generative artificial intelligence (Gen AI) across all academic disciplines (*5-8*) is recasting scientific production. Despite growing excitement (and concern) about Gen AI's role in research, empirical evidence remains fragmented, and systematic understanding of the impact of Large Language Models (LLMs) across scientific domains is limited.

Researchers have demonstrated the value of AI in many specific scientific contexts (*9-11*), such as protein structure prediction (*12*) and materials discovery (*13*). Recent advancements in LLMs have expanded its use across a wide range of tasks in natural (*14-16*) and social sciences (*17-20*). This work highlights the incredible potential of LLMs across specific scientific undertakings, raising an open question: What is the macro level impact of LLMs on the scientific enterprise?

To address this question, we collected large-scale data from three preprint repositories (Jan 2018 to June 2024, see SM S1.1-1.3 for details): (1) arXiv (1.2M preprints), which includes mathematics, physics, computer science, electrical engineering, quantitative biology, statistics, and economics, (2) bioRxiv (221K preprints), which spans a wide range of subfields in biology and the life sciences, and (3) Social Science Research Network (SSRN, 676K preprints), a working paper repository hosting manuscripts in the social sciences, law and the humanities. Each of the three datasets represents the largest within its domain. Collectively, they offer an unprecedented empirical basis to examine some of the impacts of LLMs on scientific productivity practices across many scientific fields.

To identify the use of LLMs in the creation of scientific manuscripts, we developed a text-based AI detection algorithm (*5*), which we applied to all abstracts in our data. We used abstracts from papers submitted prior to 2023 – before the ChatGPT era – to estimate the token distribution of human-written text, then prompted OpenAI's GPT-3.5turbo0125 model to rewrite these abstracts to generate the token (word) distribution of LLM-written text. We compared these distributions to identify probable LLM-assisted abstracts written after the release of ChatGPT 3.5. Further details on model training, validation, potential limitations, and alternative methods of LLM detection are provided in SM S2.1, S4, and S5.

**Results**

***LLM usage and scientific productivity***

We predicted that authors who adopted LLMs would experience increased productivity (*21-23*). To isolate the general productivity effects of LLMs from rapid growth in research on AI, we first exclude manuscripts in core AI subdisciplines (SM S1.1, S5.7) from our sample. We then identify an author's initial adoption of LLMs as marked by the first manuscript ($m_i$) that exhibited statistical signatures of LLM assistance ($\alpha$), such that $\alpha(m_j) > \tau$, where $\tau$ is the detection threshold. An author's adoption status changes from 0 to 1 for all months subsequent to detected use. Based on this measure, we examine the change in manuscript submission rates between LLM adopters and similar non-adopters (SM S2.2-2.4) before and after adoption in author-level fixed effects event models (SM S3.1).

Fig. 1 shows that LLM adoption is associated with a large increase in researchers' scientific output in all three preprint repositories. The effect sizes for arXiv, bioRxiv, and SSRN are 36.2%, 52.9%, and 59.8%, respectively, suggesting that LLM use is associated with sizable increases in productivity. Although estimated coefficients vary by detection method and threshold used to identify LLM adoption, sensitivity analyses (SM S5.3-5.6) demonstrate that a positive association is robust across analytical choices.

A productivity jump may stem from the use of Gen AI across multiple research tasks, including idea generation, literature discovery, coding, data collection or analysis. But to date, LLMs likely have had the largest impact in writing (*24, 25*). To create distinctive scientific works, researchers must present compelling written arguments; link a manuscript's arguments, methods, and results to prior literature; detail and contextualize the most important findings; and articulate what can be learned from the text. These complex writing tasks are time consuming, particularly for researchers communicating in a non-native language. We therefore ask: Does the productivity impact of LLM adoption vary across authors' native language proficiencies? Since most high-impact research is published in English-language journals and proceedings, native speakers have had a substantial advantage in scientific communication (*26-28*). LLMs can mitigate disparities in English fluency, which should asymmetrically reduce the cost of writing across scientists' linguistic backgrounds (*29*).

To test for heterogeneity in treatment effects, we approximate the likelihood that an author is a native English speaker based on names and the institutions with which they are affiliated (SM S2.5-2.6). Fig. 2 shows coefficients broken out by researchers' ethnicities and home geographies. The effects remain statistically significant across all groups, but scholars with Asian names experience the greatest productivity boost from LLM adoption. In bioRxiv and SSRN, effects are even more pronounced for scholars with Asian names and institutional affiliations in Asia, but the difference with Asian scholars in US/UK/CA/AU institutions is only statistically significant in bioRxiv. The estimated effect of LLM adoption on scientific output for this group of scholars ranges from a low of 43.0% in arXiv to 89.3% for bioRxiv and 88.9% for SSRN. Researchers with Caucasian names affiliated with institutions in English-speaking countries experience more modest but still significant productivity gains of 23.7% (arXiv) to 46.2% (SSRN).

We conclude that even the use of previous-generation LLMs — those available to scholars at the time the manuscripts in our data were drafted — are associated with productivity gains, particularly for researchers facing higher costs of writing. These findings concur with work showing that LLMs mitigate the impact of skill disparities, in this case by reducing the cost of writing in a second language (*30*). Given considerable advances in the writing ability of current-generation LLMs and more widespread availability of these systems, the productivity effects we estimate are likely substantial enough to imply a shift in the market share of scientific production toward scholars in non-native English-speaking geographies.

### *LLM use, scientific writing, and publication outcomes*

LLMs are likely to reshape science production beyond productivity effects. High-quality writing is often construed as a signal of scientific merit. Papers with clear but complex language are perceived to be stronger and are cited more frequently (*31*). Because novel scientific advances are the product of years of knowledge refinement, the ability to precisely articulate scientific discoveries is a (very imperfect) proxy for the care of a scientific team's work. The fact that LLMs can almost effortlessly produce polished, professional text describing any scientific topic raises an important question: Does LLM use reveal or conceal the quality of the underlying research?

To assess this question, we investigated how writing complexity relates to research quality and whether LLM adoption changes the signaling power of writing complexity in scientific communication. We gauge writing complexity with the additive inverse of the Flesch Reading Ease Score (SM S2.7). This measure quantifies text complexity as a composite of average sentence length and syllables per word, with higher scores indicating more complex text. As a proxy for quality, we then created a binary outcome defined as publication in a peer reviewed journal or conference by the end of our observation window (June 2024) for all preprints since 2023 (SM S2.8, S5.9).

When we correlate the additive inverse of the Flesch score with publication outcomes, three patterns emerge. First, writing complexity scores in LLM-assisted manuscripts are significantly higher compared to papers written in natural language in all three archives ($P < 0.001$, all repositories, two-tailed *t*-test) (Fig. 3A-C). This underscores the remarkable capability of LLMs to produce complex scientific writing (*23*). Second, in *non-LLM-assisted* papers across all three repositories, writing complexity is positively associated with manuscript quality as approximated by the probability of publication in a peer-reviewed venue (logistic regressions, Fig. 3E-G). These results confirm prior research showing a positive association between writing complexity and scientific merit (*32*). Third, and critically, we find a *reversal* in the relationship between writing complexity and peer-review outcomes for LLM-assisted manuscripts. For these documents, increases in writing complexity are associated with lower peer assessments of scientific merit (Fig. 3E-G).

To assess the robustness of these findings, we examined additional lexical, syntactic, and morphological features of writing (SM S5.10). We replicated the findings using lexical complexity (syllables per word) and morphological complexity (fraction of present participial clauses). Both showed the same reversal pattern in which increased writing complexity correlates negatively with publication success in LLM-assisted papers but positively in human-written papers. We also found the same pattern for the use of promotional language, further confirming that LLM adoption erodes traditional quality signals across multiple linguistic dimensions (*33*).

Myriad factors influence the publication outcomes of preprints. We cannot rule out all confounding factors, but the results remain consistent after controlling for preprint month and field of study (SM S3.2, S5.9). As a robustness check, we collected and analyzed an independent dataset from the International Conference on Learning Representation (ICLR-2024), a leading conference in machine learning (*34*). ICLR-2024 provides access to 28K referee reports for the full set of 7,243 submissions to the conference, regardless of their final acceptance status (SM S1.4). Using the peer review score assigned by experts as an alternative measure of scientific merit, Fig. 3D and 3H replicate the key findings with remarkable consistency.

The sharp contrast in quality assessments across the distribution of language complexity in the two groups — human-written and LLM-assisted manuscripts — confirms that complex LLM-generated language often disguises weak scientific contributions (*35*). These findings demonstrate the rapid erosion of a traditional heuristic. For LLM-assisted manuscripts, the positive correlation between linguistic complexity and scientific merit not only disappears; it inverts. As the effort required to produce polished prose declines, so too does its utility as a signal of an author's command of a topic (*36*). This creates a risk for the scientific enterprise, as a deluge of superficially convincing but scientifically underwhelming research could saturate the literature. If this occurs, it will bury some important ideas and force the community to waste valuable time separating genuine insights from a morass of unimportant work.

For peer reviewers and journal editors, this represents a significant issue. As a shortcut to (imperfectly) screen scientific research, writing characteristics are fast becoming uninformative signals just as the quantity of scientific communication surges. As traditional heuristics break down, editors and reviewers may increasingly rely on status markers such as author pedigree and institutional affiliation as signals of quality, ironically counteracting LLMs' democratizing effects on scientific production. One potential response is to leverage the same technology to assist in evaluating manuscripts. Specialized "reviewer agents" could flag methodological inconsistencies, verify claims, and even assess novelty. Whether this scalable approach will help editors and reviewers focus on substance over surface-level signals or introduce new and unforeseen challenges to the scientific process is a critical uncertainty.

### LLM use and Citation Behavior

Writing scientific papers also involves embedding claims and findings within existing literature. Because LLMs have the capacity to ingest and synthesize vast quantities of information, LLMs may broaden researchers' exposure to prior work (*37, 38*). Or, as some have speculated, training data may overrepresent high impact works, leading LLMs to amplify exposure to easily discoverable research (*39*). We therefore ask: How do LLMs affect the discovery of prior literature?

To evaluate these competing hypotheses, we leverage a unique dataset capturing 246M views and downloads of arXiv papers (SM S1.5), each connected with a user id, arXiv document id, and referral source (e.g., Bing or Google). This dataset allows us to explore changes in user-level reading behavior following the February 2023 release of Bing Chat (GPT-4), the first widely adopted AI-powered search tool (SM S3.4). Figure 4A-C compares arXiv documents accessed by Bing users before and after this exogenous shift. Our estimations based on a differences-in-differences analysis show that, compared to accesses redirected by Google, post-Bing Chat users access a more diverse set of arXiv documents. Fig. 4A compares publication formats, showing that Bing users access books at a 26.3% higher rate ($P < 0.001$, Poisson regression), presumably reflecting an LLM's ability to surface content embedded in lengthy texts.

Increased access to books suggest that LLM-aided science may draw on a broader range of reference materials, but it does not rule out that LLMs simply reinforce attention to scientific canons. We investigate this possibility and find that Bing-referred visits were also linked to more recent scholarship; the median age of manuscripts accessed decreased by an estimated 0.18 years (Fig. 4B). Furthermore, LLM users do not increase citations to well-cited works (Fig. 4C). Instead, we find that Bing users uncovered references with fewer existing citations.

To examine whether this shift in search results translated to a change in actual citation behavior, we linked preprints in arXiv, bioRxiv, and SSRN to two large-scale citation databases, OpenAlex and Semantic Scholar. We obtained 101.6M citations to prior works (SM S1.6). We then used the event study methodology in Fig. 1 to compare authors' citation behavior before and after they adopt LLMs, relative to a control group of non-users (SM S3.3). Our analysis explores three characteristics of cited references: (1) publication format (citations to books), (2) time lag (median reference age), and (3) impact of cited work (mean log citations of referenced documents).

Consistent with Fig. 4A-C, we find that LLM use alters authors' citation behavior, seemingly steering them toward a more diverse knowledge base.. We find that LLM adopters overall are 11.9% more likely to cite books (Fig. 4D), but the effect is not statistically significant in one of the archives, SSRN. Adopters also cite documents that are on average 0.379 years younger (Fig. 4E) and show no evidence of being more impactful (2.34% lower citation impact, Fig. 4F). Although the magnitude of these effects varies by preprint repository, the overall pattern is broadly consistent (Fig. 4G–I).

Fig. 4 presents consistent evidence that AI assistance directs scholars to a broader body of knowledge. Researchers face time and attention constraints that limit their ability to process the expanding universe of research (*40, 41*). LLMs appear to help researchers overcome obstacles in discovering pertinent literature.

These findings suggest that while LLMs may obscure signals of authorial effort, they appear to clarify the path to knowledge discovery. A common concern has been that AI-assisted search might reinforce the existing scientific canons. We find, however, that LLM adoption has had the opposite effect. Both AI-assisted search behavior and author citation patterns show a significant shift toward a more diverse knowledge base, one that includes more books as well as younger and less-cited scholarship. This broadening of attention suggests that LLMs help researchers overcome cognitive constraints that have limited their ability to engage with the ever-expanding universe of scientific literature.

### Limitations and Future Directions

This study explores the impact of LLMs on scientific production, but our findings are subject to several limitations that offer avenues for future research.

First, we do not provide causal identification. A central difficulty in studying LLMs "in the wild" is the impossibility of perfectly measuring their use. Our AI detection method is imperfect and susceptible to several challenges (SM S5.1-5.4): it relies on abstracts rather than full text (SM S5.5); it cannot definitively identify which specific co-author on a team used an LLM (SM S5.6); and it almost certainly fails to detect use by authors who heavily edit LLM-assisted text. Furthermore, the non-random adoption of Gen AI tools creates the potential for self-selection bias, and our focus on published papers means the "adoption time" may be endogenous to productivity. Our supplement contains many additional analyses to evaluate the scope of these issues, and while our results appear robust, it is important for future work to continue to identify methodological strategies to address these challenges.

Second, our findings represent a snapshot of a rapidly evolving technology. Our analysis is based on data generated prior to the arrival of more advanced reasoning models and deep research capabilities. As models improve and scientists discover new ways to integrate them into their work, the future impact of these technologies will likely dwarf the effects we have highlighted here. This presents a crucial direction for future research: to continuously track how the scientific enterprise incorporates successive generations of AI models. Studies will need to examine whether the effects we document are amplified, altered, or even reversed as these more powerful tools are integrated into the scientific workflow.

There are many directions for future research. A primary avenue is more nuanced explorations of how LLMs are impacting scientific practice. Advancement in science has long been constrained by access to informal resources and knowledge — what Diana Crane termed "invisible colleges" (*42*). One hypothesis is that LLMs provide a scalable substitute for this informal knowledge, offering guidance on everything from experimental design to navigating a field's hidden curriculum, thereby leveling the scientific playing field. Another interesting avenue for future research is the potential for LLMs to transcend disciplinary boundaries. Over time, academic disciplines have developed deep knowledge bases that are often communicated through discipline-specific jargon. If LLMs help outsiders to overcome this hurdle, siloed disciplines may more productively engage with one another.

**Conclusion**

Our findings show that LLMs have begun to reshape scientific production. Use of LLMs accelerates manuscript output, reduces barriers for non-native English speakers, and diversifies the discovery of prior literatures. These changes may democratize scientific production. However, traditional signals of scientific quality such as language complexity are becoming unreliable indicators of merit just as we experience an upswing in the quantity of scientific work. These changes portend an evolving research landscape in which the value of English fluency will recede, but the significance of robust quality assessment frameworks and deep methodological scrutiny is paramount. As AI systems advance, they will challenge our fundamental assumptions about research quality, scholarly communication, and the nature of intellectual labor. An urgent agenda for science policy makers will be to evolve our scientific institutions to accommodate the rapidly evolving scientific production process.

**Acknowledgments:** We thank Mor Naaman, Will Cong, Wu Zhu, as well as seminar participants at Complexity Science Hub (Vienna), UCLA Price Center, Haas MRL seminar, and Columbia MAD conference for helpful discussions, and Alex Cui for providing academic access to the GPTZero API. This work is supported by National Science Foundation under grant numbers 2311521, 2404035, and 2412389.

**Supplementary Materials**

Supplementary Text

Figs. S1 to S37

Tables S1 to S10

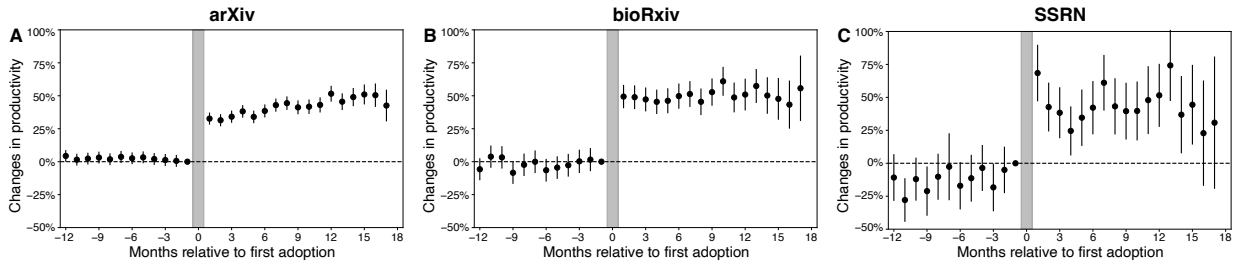

**Figure 1. LLM usage and scientific productivity.** We track the productivity dynamics (measured as the number of preprints published monthly) of 168,553 authors between Jan 2022 and July 2024, distributed across arXiv (109,965 authors), bioRxiv (43,218 authors), and SSRN (15,370 authors). For each author, we apply a text-based detector to their preprints to determine whether and when they "adopted" LLMs in scientific writing. **(A-C)** Using a stacked difference-in-difference regression, we estimate the impact of LLM on individual productivity. Comparing researcher-level pre- and post-adoption, we conservatively observe significant productivity increases, with boosts of 36.2% (arXiv), 52.9% (bioRxiv), and 59.8% (SSRN) relative to non-adopters.

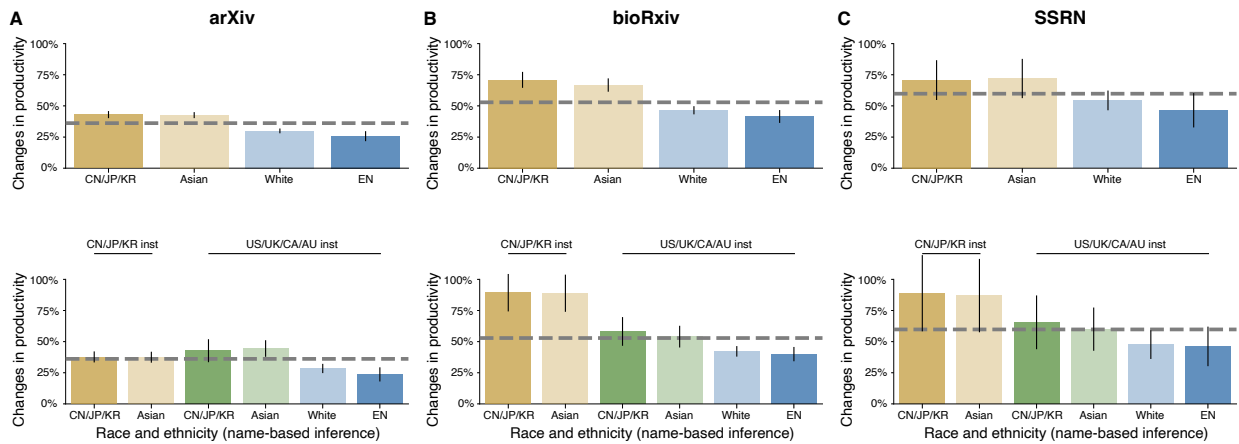

**Figure 2. Heterogeneity across races, ethnicities and home geographies.** (**A-C**) The productivity gains are pronounced for authors with East Asian names, showing increases of 43.0% (arXiv), 70.9% (bioRxiv), and 70.1% (SSRN), but remain meaningful for authors with Caucasian names, with gains of 25.7% (arXiv), 41.5% (bioRxiv), and 46.5% (SSRN). The effect is most pronounced for authors with *Asian names affiliated with institutions in Asia*, showing productivity boosts of 37.7% (arXiv), 89.3% (bioRxiv), and 88.9% (SSRN). By comparison, the productivity boost for authors with Caucasian names affiliated with institutions in English-speaking countries, are 23.7% (arXiv), 40.0% (bioRxiv), and 46.2% (SSRN). The gray dashed line represents the average effect across all authors.

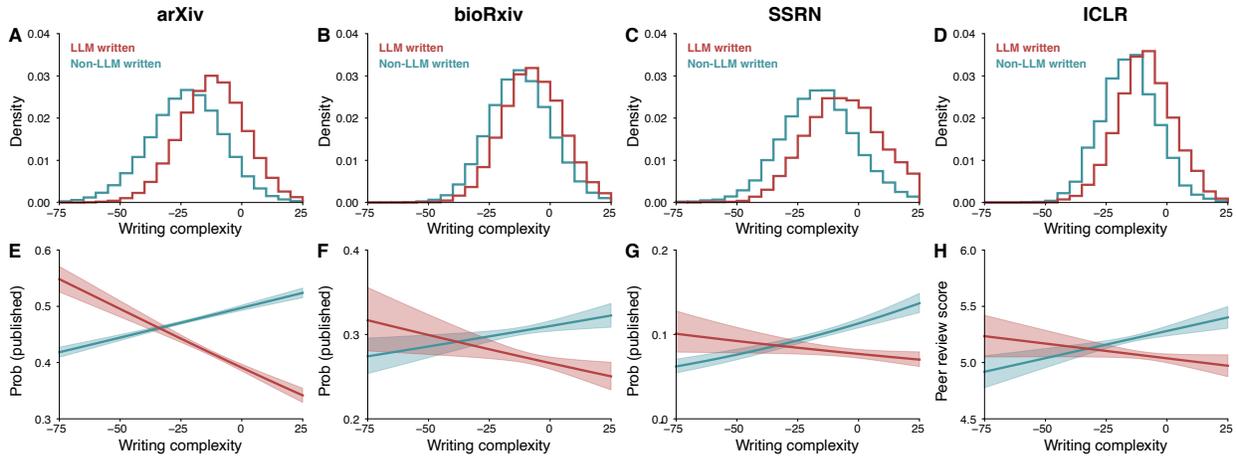

**Figure 3. LLM usage, scientific writing, and publication outcomes.** For 264,125 manuscripts written since 2023 (177,880 on arXiv, 58,958 on bioRxiv, 31,959 on SSRN, and 7,243 submitted to ICLR 2024). We measure writing complexity as the negative of the Flesch Reading Ease Score, a metric defined as a joint function of a text's mean sentence and word length. We also measure paper quality using (i) the probability of peer-reviewed publication for preprints (prior to the data cutoff), and (ii) peer review scores for ICLR submissions. **(A-D)** Distribution of writing complexity for LLM-assisted (red) and non-LLM-assisted (blue) manuscripts. LLM-assisted manuscripts exhibit significantly higher writing complexity across all four datasets. **(E-H)** Relationship between writing complexity and paper quality. For non-LLM-assisted manuscripts, writing complexity is positively correlated with measures of perceived manuscript quality. In sharp contrast, for LLM-assisted manuscripts, greater writing complexity correlates with lower manuscript quality. Predicted outcomes are based on logistic regressions (publication probability in **E-G**) and OLS regressions (peer review scores in **H**), respectively.

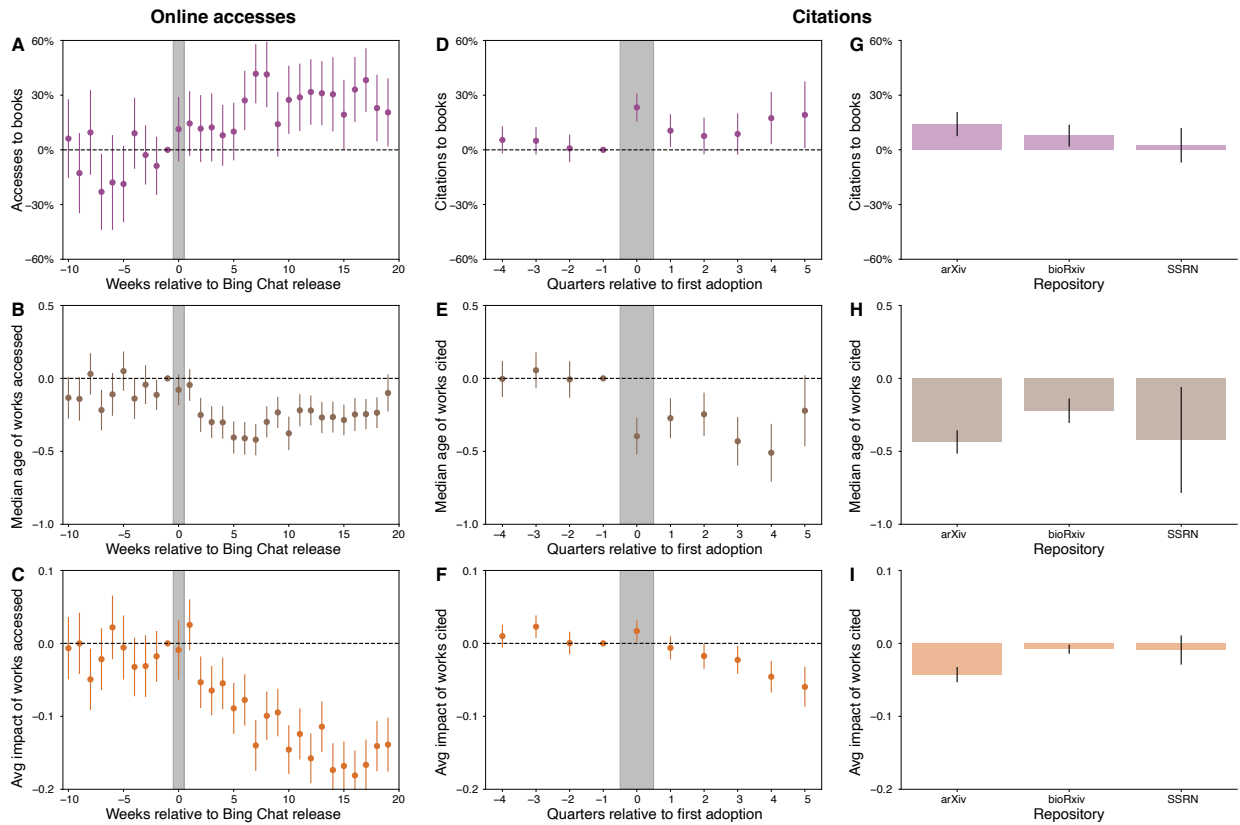

**Figure 4. LLM usage and references to prior works. (A-C)** We examine user-level changes in access to arXiv manuscripts following the release of Bing Chat (powered by GPT-4) in Feb 2023. Comparing online accesses redirected from Google and Bing, we find users access **(A)** more books**, (B)** more recent works, and **(C)** less highly cited works, post-event. **(D-I)** We examine author-level changes in referencing / citation patterns following authors' initial adoption of LLMs, using a Diff-in-Diff strategy as in Figure 1. Due to data sparsity, all outcomes are measured quarterly. **(D)** Authors using LLMs cite 11.9% more books post-adoption, showcasing LLM's advanced ability to process and integrate a more diverse range of knowledge sources. Estimates are from Poisson regressions. **(E)** Authors using LLMs cite more recent references post-adoption, with the median reference age decreasing by 0.379 years. **(F)** Contrary to concerns that LLMs may reinforce reliance on well-established scientific works, we find no increase in citation to high-impact works, where "reference impact" is measured by log(#citations+1) averaged over all cited references. **(G)** Authors using LLMs on arXiv, bioRxiv, and SSRN cite 14.1% ($P < 0.001$), 7.81% ($P = 0.011$), and 2.50% ($P = 0.605$) more books post-adoption, respectively. **(H)** Authors using LLMs on arXiv, bioRxiv, and SSRN cite references that are 0.436 years ($P < 0.001$), 0.222 years ($P < 0.001$), and 0.422 years ($P = 0.023$) more recent, respectively. **(I)** Authors using LLMs on arXiv, bioRxiv, and SSRN cite references that are 4.29% ($P < 0.001$), 0.800% ($P = 0.012$), and 0.916% ($P = 0.369$) less highly cited, respectively.